\documentclass[]{aa}
\usepackage[utf8]{inputenc}
\usepackage{amsfonts}       
\usepackage{graphicx}
\usepackage{natbib}
\usepackage{textcomp}
\usepackage{physics}
\newcommand{\RomanNumeralCaps}[1]
    {\MakeUppercase{\romannumeral #1}}

\title{Swirls in the Solar Corona}

\author{C.~Breu\inst{\ref{inst1}\and \ref{inst2}}\and H.~Peter\inst{\ref{inst1}}\and R.~Cameron\inst{\ref{inst1}}\and S.K.~Solanki\inst{\ref{inst1}\and \ref{inst3}}}
\authorrunning{Breu et al.}
\institute{Max Planck Institute for Solar System Research, Justus-von-Liebig-Weg 3, 37077 Göttingen, Germany\label{inst1} \and School of Mathematics and Statistics, University of St. Andrews, St. Andrews, Fife, KY16 9SS, UK\label{inst2} 
\and School of Space Research, Kyung Hee University, Yongin, Gyeonggi 446-701, Republic of Korea\label{inst3}
}

\begin{document}

\abstract{Vortex flows have been found in the photosphere, chromosphere and low corona in observations and simulations. It has been suggested that vortices play an important role for channeling energy and plasma into the corona, but the impact of vortex flows on the corona has not directly been studied in a realistic setup.}
{We investigate the role vortices play for coronal heating using high resolution simulations of coronal loops. The vortices are not artificially driven, but arise self-consistently from magnetoconvection.}
{We perform 3D resistive MHD simulations with the MURaM code. Studying an isolated coronal loop in a Cartesian geometry allows us to resolve the structure of the loop interior. We conduct a statistical analysis to determine vortex properties as a function of height from the chromosphere into the corona.}   {We find that the energy injected into the loop is generated by internal coherent motions within strong magnetic elements.
A significant part of the resulting Poynting flux is channeled through the chromosphere in vortex tubes forming a magnetic connection between the photosphere and corona. Vortices can form contiguous structures that reach up to coronal heights, but in the corona itself the vortex tubes get deformed and eventually lose their identity with increasing height.  Vortices show increased upward directed Poynting flux and heating rate both in the chromosphere and corona, but their effect becomes less pronounced with increasing height.}{While vortices play an important role for the energy transport and structuring in the chromosphere and low corona, their importance higher up in the atmosphere is less clear since the swirls are less distinguishable from their environment. Vortex tubes reaching the corona show a complex relationship with the coronal emission.}

\maketitle

\section{Introduction}

Vortex flows driven by magnetoconvection have been proposed as a possible energy channel from the convection zone to the upper solar atmosphere \citep{2012Natur.486..505W}.
Rotating structures have been detected in the photosphere and chromosphere in high resolution observations \citep{1988Natur.335..238B, 2008ApJ...687L.131B,2010ApJ...723L.139B,2009A&A...507L...9W} and simulations \citep{2010MmSAI..81..582C,2011A&A...526A...5S,2011A&A...533A.126M,2011ApJ...727L..50K,2012A&A...541A..68M,2012ASPC..456....3S,2012PhyS...86a8403K,2012ApJ...751L..21K,2020ApJ...894L..17Y,2021A&A...645A...3Y}. A comprehensive overview over observations of vortical flows in different layers of the solar atmosphere is given by \citet{2023SSRv..219....1T}. Recently, vortices have been investigated in a coronal hole simulation \citep{2022A&A...665A.118F}. In addition to predominantly vertical vortices, \citet{2010ApJ...723L.180S} identified horizontal vortices along the edges of granules by comparing observations with simulations.\\
The photospheric vortices found in numerical simulations have sizes of the order of the width of the downflow lanes or smaller \citep{2011A&A...533A.126M}.
It is not yet known what drives the formation of these small-scale vortices. 
\citet{1998ApJ...499..914S} find regions of enhanced vorticity in downflow regions in the intergranular lanes and suggest that vorticity is produced at granule edges and in downdrafts by a misalignment of pressure and density gradients and then concentrated in the intergranular lanes by advection.  The conservation of the angular momentum of the inflows results in the bathtub effect, amplifying the rotational motion in the narrow downflow lanes \citep{1985SoPh..100..209N}. 
In contrast to this, \citet{2012PhyS...86a8403K} propose that instead vorticity is generated by convective granular instability or the Kelvin-Helmholtz instability developing between shearing flows in the intergranular lanes.\\
In a plasma with a high magnetic Reynolds number such as the solar atmosphere, magnetic field lines are frozen into the plasma. Furthermore, in the high plasma beta environment of the photosphere, the evolution of the magnetic field is determined by plasma flows. The rotating downflows wind up the magnetic field in the intergranular lanes. The twisting of the magnetic field has been proposed to help to confine magnetic flux tubes by the pinch effect \citep{1975SoPh...42...79S}, although this is now not considered to be of any real importance, with the main confinement being produced by the evacuation of magnetic flux tubes (\citealp{1976SoPh...50..269S}, \citealp[cf.][]{1993SSRv...63....1S}).
In the chromosphere, swirls have been observed as dark structures in the Ca \RomanNumeralCaps{2} 8542 \AA\ line \citep{2009A&A...507L...9W}.
Multiwavelength observations of chromospheric swirls co-located with magnetic bright points have shown that vortices reach from the photosphere into the low corona \citep{2009A&A...507L...9W,2012Natur.486..505W}. An imprint of rapidly rotating magnetic structures has been detected in transition region and coronal emission in the 304, 171 \AA\ and 211 \AA\ channels of AIA corresponding to hot coronal plasma \citep{2012Natur.486..505W}. Only four out of 14 detected swirls, however, show coronal counterparts. A persistent vortex flow reaching the low corona was also reported by \citet{2018A&A...618A..51T}. In contrast to \citet{2012Natur.486..505W}, \citet{2018A&A...618A..51T} find a decrease in intensity in the 171 \AA\ channel and no signal in the 193 or 211 \AA\ channels. The vortices are also seen as absorbing structures in \citet{2012Natur.486..505W}, but the authors noted higher intensities in some AIA channels at the edges of the vortex flows. The number of chromospheric swirls has been estimated as $2\times 10^{-3}\; \rm{Mm^{-2}}$ by \citet{2012Natur.486..505W}, $6.1\times 10^{-2}\; \rm{Mm^{-2}}$ by \citet{2019NatCo..10.3504L} and $8\times 10^{-2}\; \rm{Mm^{-2}}$ by \citet{2022A&A...663A..94D}. The diameter of the observed chromospheric swirls ranges from 0.58 Mm \citep{2019NatCo..10.3504L} to 4.4 Mm \citep{2018A&A...618A..51T}. Measured average lifetimes in observations lie between 21 s \citep{2019NatCo..10.3504L} and more than 1.7 h \citep{2018A&A...618A..51T}. For a comprehensive overview of observed properties of swirls see the review by \citet{2023SSRv..219....1T}.\\
The observed enhanced emission in coronal spectral lines points to heating of the swirls.
Heating at the location of vortices has been found in simulations in the upper photospheric layers due to viscosity \citep{2012A&A...541A..68M} and in the chromosphere, for example in \citet{2012ApJ...751L..21K,2020ApJ...894L..17Y,2021A&A...645A...3Y}.
The main contribution to the Poynting flux in the atmosphere was suggested to stem from horizontal motions acting on the strong predominantly vertical magnetic field \citep{2012ApJ...753L..22S}. 
\citet{2020ApJ...894L..17Y} found that vortices contribute a significant fraction of the Poynting flux in the chromosphere.
An increased Poynting flux associated with vortex locations was also reported by \citet{2021A&A...649A.121B}.
Vortex motions could thus be partially responsible for heating the solar chromosphere.\\\\
In addition to acting as an energy channel, vortices could also be associated with the transport of matter.
\citet{2009A&A...507L...9W} found that the chromospheric swirls were blue-shifted, which they interpreted as plasma moving upward along a twisted magnetic field. \citet{2018A&A...618A..51T} also detected upflows within the observed swirling structure. Vortices could be related to the ejection of jets and spicules which have been observed to be spinning \citep{2009A&A...507L...9W,2012ApJ...752L..12D}.
\citet{2012ApJ...751L..21K,2013ApJ...770...37K} find that the vortices in their simulation are associated with quasi-periodic upflows at the vortex edge driven by the Lorentz force arising from current sheets in the atmosphere and by the pressure gradient in the subsurface layers, while at the same time a helical downflow is present in the vortex core. \citet{2017ApJ...848...38I} showed that the twisted magnetic field in a vertical vortex can drive chromospheric jets via the Lorentz force.\\\\
Vortices could launch Alfv\'{e}n waves that propagate into the corona.
\citet{2013ApJ...776L...4S} interpret vortices not as continuous rotational flows, but instead as torsional Alfv\'{e}n waves propagating along the guide field excited by oscillatory motions of the plasma in magnetic concentrations. This view has been challenged by \citet{2021A&A...649A.121B}, who instead find unidirectional upward propagating pulses that could contribute to chromospheric and coronal heating. Various wave modes in a long-lived vortex structure have been observed by \citet{2020A&A...643A.166T}. \citet{2022A&A...665A.118F} suggest that vortices could be responsible for heating and acceleration of coronal and solar wind plasma.\\
Common to most existing simulations is that they have an upper boundary located in the chromosphere, transition region or low corona. The first mention of vortex flows located in downdrafts in solar surface convection simulations occurs in \citet{1985SoPh..100..209N}. The computational domain of this simulation has a vertical extent of 1.6 Mm. \citet{2011A&A...533A.126M} reach heights of 500 to 600 km above the $\langle\tau\rangle=1$ layer, the simulations studied in \citet{2012A&A...541A..68M} and \citet{2013ApJ...776L...4S} reach 800 km above the photosphere. The $\rm{CO^{5}BOLD}$ simulation used by \citet{2012Natur.486..505W} has an upper boundary located in the upper chromosphere at a height of 2 Mm, while the Bifrost snapshot studied in the same article has a vertical extent of 15.4 Mm.
The simulation analyzed in \citet{2020ApJ...894L..17Y,2021A&A...645A...3Y} has an upper boundary set in the upper chromosphere at 2.5 Mm. \citet{2013ApJ...770...37K} used an upper boundary in the chromosphere that extends to 1 Mm above the photosphere. Studies including a coronal part are the simulation by \citet{2017ApJ...848...38I} with an upper boundary at a height of 8 Mm in an open field setup and the coronal hole setup by \citet{2022A&A...665A.118F}. The first study, however, did not investigate the Poynting flux transport and dissipation inside the structure. \citet{2022A&A...665A.118F} find enhancements in temperature and density that they attribute to the vortices. An extensive overview of simulations including atmospheric vortices can be found in \citet{2023SSRv..219....1T}.\\
The choice of the top boundary conditions can affect the behavior of the thermodynamic quantities, for example an increased diffusivity near the boundary \citep{2012A&A...541A..68M,2020ApJ...894L..17Y,2021A&A...645A...3Y}.\\
In this study, we investigate the influence of vortices on heating and structuring of the corona in a stretched-loop setup including a chromospheric layer and a shallow convection zone layer at each end of the simulation box. Most previous studies of vortices in the solar atmosphere use an open field configuration \citep{2011A&A...533A.126M,2012A&A...541A..68M,2013ApJ...776L...4S,2012Natur.486..505W,2020ApJ...894L..17Y,2021A&A...645A...3Y,2013ApJ...770...37K,2017ApJ...848...38I,2022A&A...665A.118F}, apart from the Bifrost snapshot used in \citet{2012Natur.486..505W}, which contains a coronal loop structure connecting regions of opposite polarity. In this snapshot, a swirl is found at one of the loop footpoints, hinting that vortices could also contribute to injecting energy into coronal loops (presented in the supplementary material of that paper). \citet{2015Natur.522..188A} consider a magnetic field generated by a small-scale dynamo operating in the subsurface layers superposed with a vertical field mimicking the footpoints of quiet-Sun magnetic loops reaching high altitudes. We use a modified version of the MURaM code to conduct full 3D MHD simulations. The modifications to the code are discussed in detail by \citet{2022A&A...658A..45B}. Vortices in our simulation are self-consistently excited by magnetoconvection and their effect on the solar atmosphere is not affected by the choice of the boundary conditions at the upper boundary.\\
The evolution of the swirling structures in higher layers of the atmosphere, and their contribution to the energy transport and the impact of the energy and mass transfer in the chromosphere has not yet been studied in detail.
Including a corona makes it possible to study the propagation of perturbations as well as the effects of heating events in higher atmospheric layers on the cooler chromospheric plasma, such as chromospheric evaporation.\\
The aim of this paper is to investigate the role the vortex flows play for the energy and mass transport between the photosphere and the corona in coronal loops and their influence on coronal loop structure.
The paper is structured as follows: Section \ref{section:meth_swirls} describes the numerical setup and analysis methods. The results are presented in Section \ref{section:results_swirls} and subsequently discussed in Section \ref{section:discussion_swirls}. Conclusions are given in Section \ref{section:conclusion_swirls}. 

\section{Methods}
\label{section:meth_swirls}

\subsection{Numerical setup}

\begin{figure*}
\resizebox{\hsize}{!}{\includegraphics{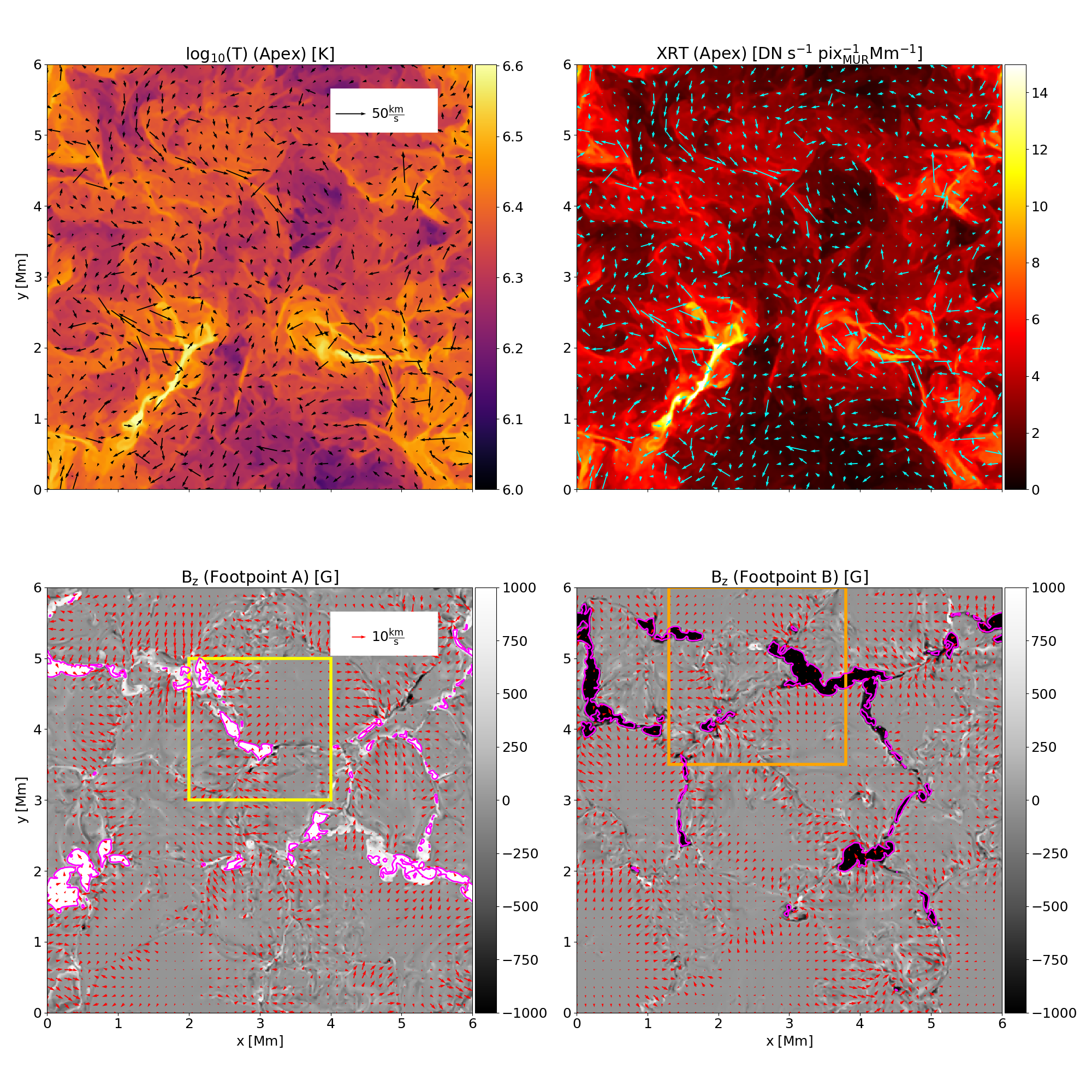}}
  \caption{Overview over the simulation box. Top row: Temperature distribution (left) and the emission as it would be seen with the Al-poly filter of the Hinode/XRT X-ray imager at the loop apex (right). Bottom row: Vertical magnetic field at the $\langle \tau \rangle =1$ layer at both loop footpoints at s=0 (left, footpoint A) and s=50 (right, footpoint B). The arrows show magnitude and direction of the velocity field perpendicular to the loop axis. The yellow and orange rectangles mark the position of the closeups of the footpoints shown in Fig. \ref{fig:swirl_fp}. The magenta contours outline magnetic field concentrations with $\vert B_{z}\vert \geq 1000\; \rm{G}$. The snapshot was taken at T=2.23 min. Data is taken after running the simulation at high resolution for 30 min. See Sect. \ref{section:meth_swirls}.}
  \label{fig:box_overview}
\end{figure*}

We model an isolated coronal loop in a Cartesian box using the MURaM code with the coronal extension \citep{2005A&A...429..335V, 2017ApJ...834...10R}. The setup is similar to the loop model described in \citet{2022A&A...658A..45B}. The coronal loop is modeled as a straightened magnetic flux tube with a shallow convection zone included at the loop footpoints at each end of the box. In the following, we will refer to the loop footpoints at each end of the simulation domain as footpoint A and footpoint B. The simulation domain includes the convection zone, the photosphere, an LTE chromosphere and a corona spanning between the footpoints. The coronal part is heated self-consistently by magnetoconvection at the loop footpoints. 
The magnetic field configuration at the footpoints and temperature as well as resulting X-ray emission are shown in Fig. \ref{fig:box_overview}. 
We solve the compressible MHD equations by conducting 3D resistive MHD simulations including the effects of field-aligned heat conduction, optically thick gray radiative transfer in the photosphere and chromosphere and optically thin losses in the corona. The treatment of the radiative transfer, heat conduction and resistivity and viscosity are described in \citet{2005A&A...429..335V} and \citet{2017ApJ...834...10R}.
To close the system of equations, we use a non-ideal equation of state.\\
We model the plasma as a single fluid and neglect effects arising from the interaction between ions and neutrals. These effects could influence the absorption of Poynting flux in the chromosphere \citep{2012ApJ...747...87K, 2016ApJ...819L..11S}. We do not take into account effects from non-local thermodynamic equilibrium in the chromosphere. While a more accurate treatment of the chromosphere will influence the temperature and electron distribution in the middle and upper chromosphere \citep{2022A&A...664A..91P}, the lower chromosphere is in LTE and the interaction between plasma and magnetic field that leads to the formation of vortices should not be significantly affected.\\
The effective loop length is 50 Mm and the photosphere is located at an average height of 3.5 Mm above the bottom boundary. The horizontal extent of the simulation box is $6\times 6\; \mathrm{Mm}$. Throughout the paper, the coordinate s refers to the distance to the approximate optical surface of the left loop footpoint along the semi-circular arclength of the loop. The approximate optical surface is defined as $\langle \tau \rangle = 1$. 
The spatial resolution is $\Delta x =12\; \mathrm{km}$ in both the horizontal and vertical direction. 
We limit the Alfv\'{e}n speed to $v_{A}=6000\; \mathrm{km\;s^{-1}}$ to avoid strict limits on the time step.\\
The initial condition for the magnetic field is a uniform vertical field of 60 G. The high resolution simulation used in this study is interpolated, in the first timestep, from a snapshot of a corresponding simulation with a lower resolution of 60 km. The simulation is run for 30 minutes with the new resolution to let initial transients subside before the results are analyzed.

\subsection{Vortex detection}

Identifying vortices in a systematic way is not straightforward. We use the swirling strength criterion by \citet{1999JFM...387..353Z} to detect vortices in the simulation. The swirling strength criterion is more reliable in the detection of vortices than an increased vorticity alone, since the vorticity is not only enhanced in rotational flows but also in shear flows without rotation  \citep{2012A&A...541A..68M}.\\
The velocity gradient tensor $U_{ij}=(\partial_{j}v_{i})$ can have all real eigenvalues, or one real eigenvalue and two complex conjugate eigenvalues \citep{1990PhFlA...2..765C}.
According to the swirling strength criterion, a grid cell is part of a vortex if the velocity gradient tensor  has a complex-conjugated pair of eigenvalues.
The eigenvector corresponding to the real eigenvalue $\lambda_{\rm{r}}$ is identified as the direction of the vortex. The swirling strength is then defined as the unsigned imaginary part $\lambda_{\rm{ci}}$ of the complex eigenvalue. The sign of the swirling strength determines whether the vortex rotates clockwise or counterclockwise.
The shear part of the vorticity can be computed as $\omega_{\rm{shear}} = \omega -2\lambda_{\rm{ci}}$.\\
Since this criterion selects the smallest vortices present in the simulation domain, we follow the approach by \citet{2020ApJ...894L..17Y} and smooth the velocity field with a Gaussian kernel with a FWHM of 500 km before applying the swirling strength criterion in order to capture larger rotating structures. 
Examples for alternative detection methods are the $\Gamma$-function criterion \citep{2001MeScT..12.1422G}, Objective Lagrangian Vortex detection (LAVD) \citep{2016JFM...795..136H} or the SWIRL algorithm \citep{2022A&A...668A.118C}. 
Comparing vortex properties depending on the vortex detection method is the subject of future work.
For a detailed description of the swirling strength criterion as well as alternative detection methods, see \citet{2023SSRv..219....1T}.

\begin{figure*}
\resizebox{\hsize}{!}{\includegraphics{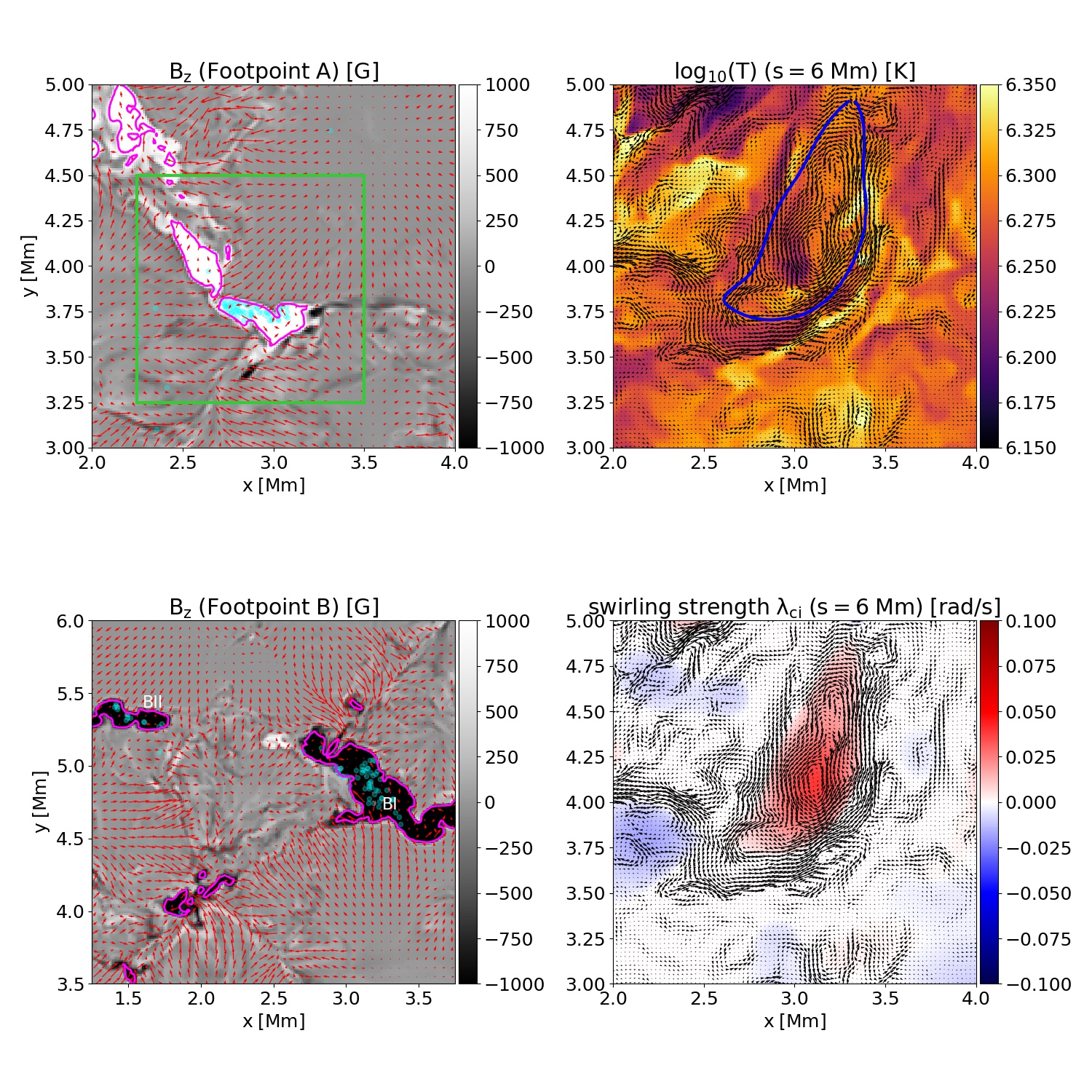}}
  \caption{Swirling strength, temperature and connection to the footpoints. \textit{Left column}: Cuts at the photospheric level showing the vertical magnetic field component at both footpoints. The cut shown in the top left panel is located at a height of 0 Mm (footpoint A), the cut shown in the bottom left panel at an arclength of 50 Mm (footpoint B). The displayed cutouts correspond to the yellow and orange squares in Fig. \ref{fig:box_overview}. \textit{Upper right panel}: Temperature at s=6 Mm. The blue contour is outlining a patch of enhanced swirling strength ($\vert\lambda_{\rm{ci}}\vert = 0.002\; \rm{rad\; s}^{-1}$ for an effective resolution of 500 km after smoothing the velocity field to bring out the larger structures). The swirling strength calculated from the smoothed flow field is shown in the bottom right panel at an arclength of 6 Mm. The light blue circles in the left column show the intersection of the magnetic field lines traced from the region outlined in blue in the upper right panel with the photospheric layer. The magenta contours outline kilogauss magnetic field concentrations. The field of view of the closeup in Fig. \ref{fig:fp_closeup} is indicated by the green square. The arrows show direction and magnitude of the velocity field perpendicular to the loop axis. See Sects. \ref{section:atmo_coup} and \ref{section:cor_heat}.}
  \label{fig:swirl_fp}
\end{figure*}
\begin{figure}
\resizebox{\hsize}{!}{\includegraphics[width=1.5\textwidth]{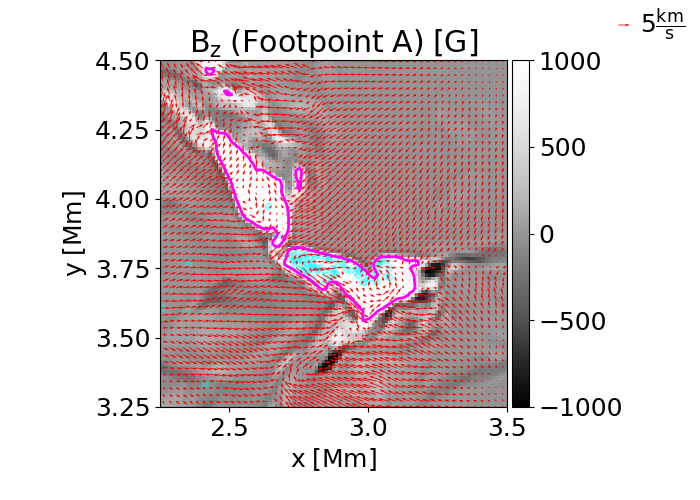}}
  \caption{Closeup of the footpoint of the swirl shown in Fig. \ref{fig:swirl_fp}. The field of view corresponds to the region within the yellow rectangle in Fig. \ref{fig:swirl_fp}. Contours in magenta mark regions with $\vert B_{z}\vert\geq 1000\;\mathrm{G}$. The red arrows show the magnitude and direction of the velocity field at the $\langle \tau \rangle =1$ surface. The light blue markers correspond to the intersections of the field lines traced from the swirl under consideration with the photospheric layer. See Sect. \ref{section:atmo_coup}.}
  \label{fig:fp_closeup}
\end{figure}

\begin{figure*}
\resizebox{\hsize}{!}{\includegraphics{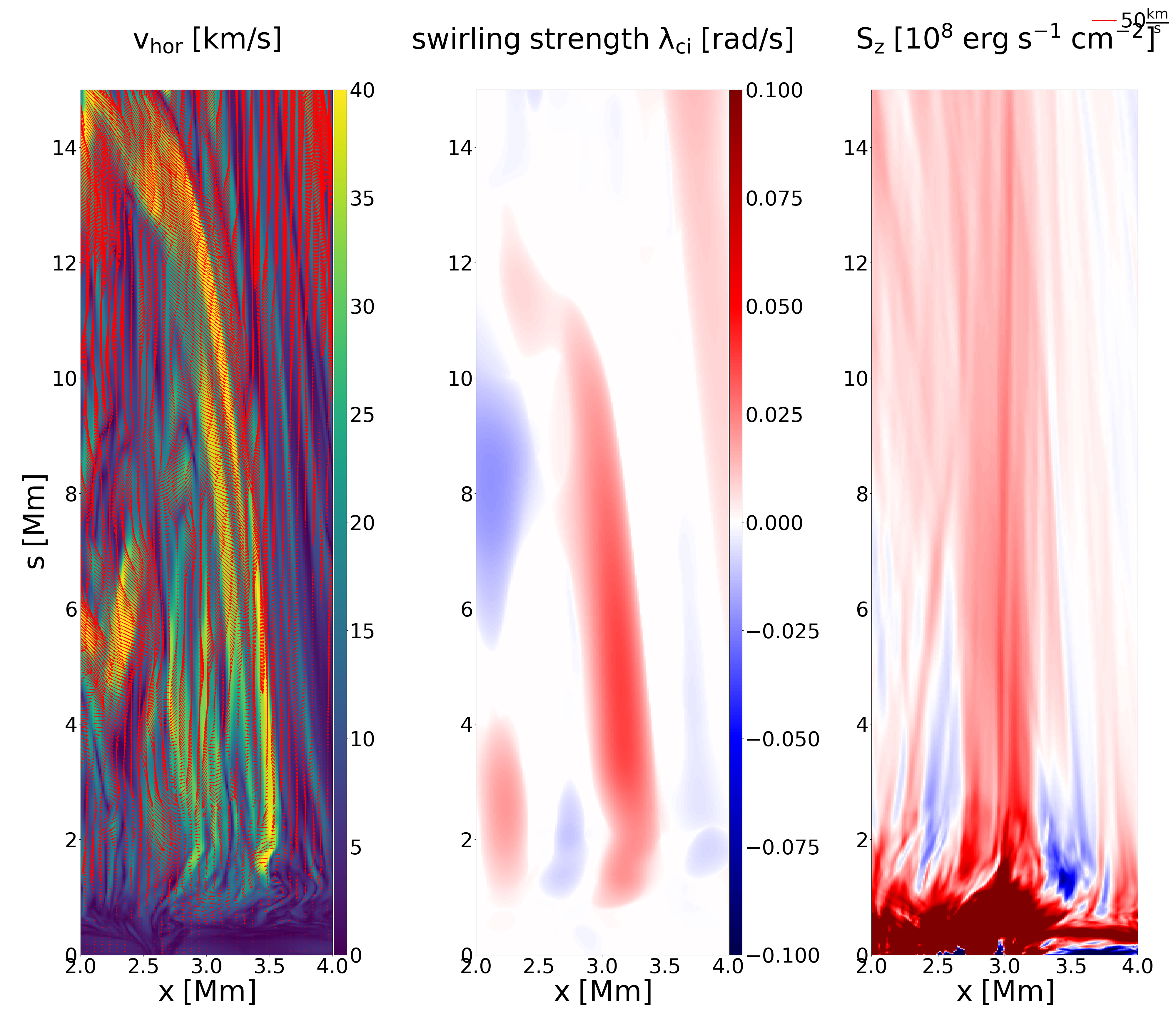}}
  \caption{Axial cut through the loop at y=4 Mm. From left to right: Velocity transverse to the guide field (color, the red arrows illustrate the velocity field projected onto the plane of the axial cut), Swirling strength computed after smoothing the velocity field with a Gaussian with an FWHM of 500 km, axial component of the Poynting flux averaged over a slab between y=3.5 Mm and y=5 Mm centered on the swirling structure. An arclength of s=15 Mm corresponds to a height above the optical surface of roughly h=12.9 Mm. For a discussion, see Sect. \ref{section:atmo_coup}.}
  \label{fig:swirl_avg}
\end{figure*}

\begin{figure*}
\resizebox{\hsize}{!}{\includegraphics{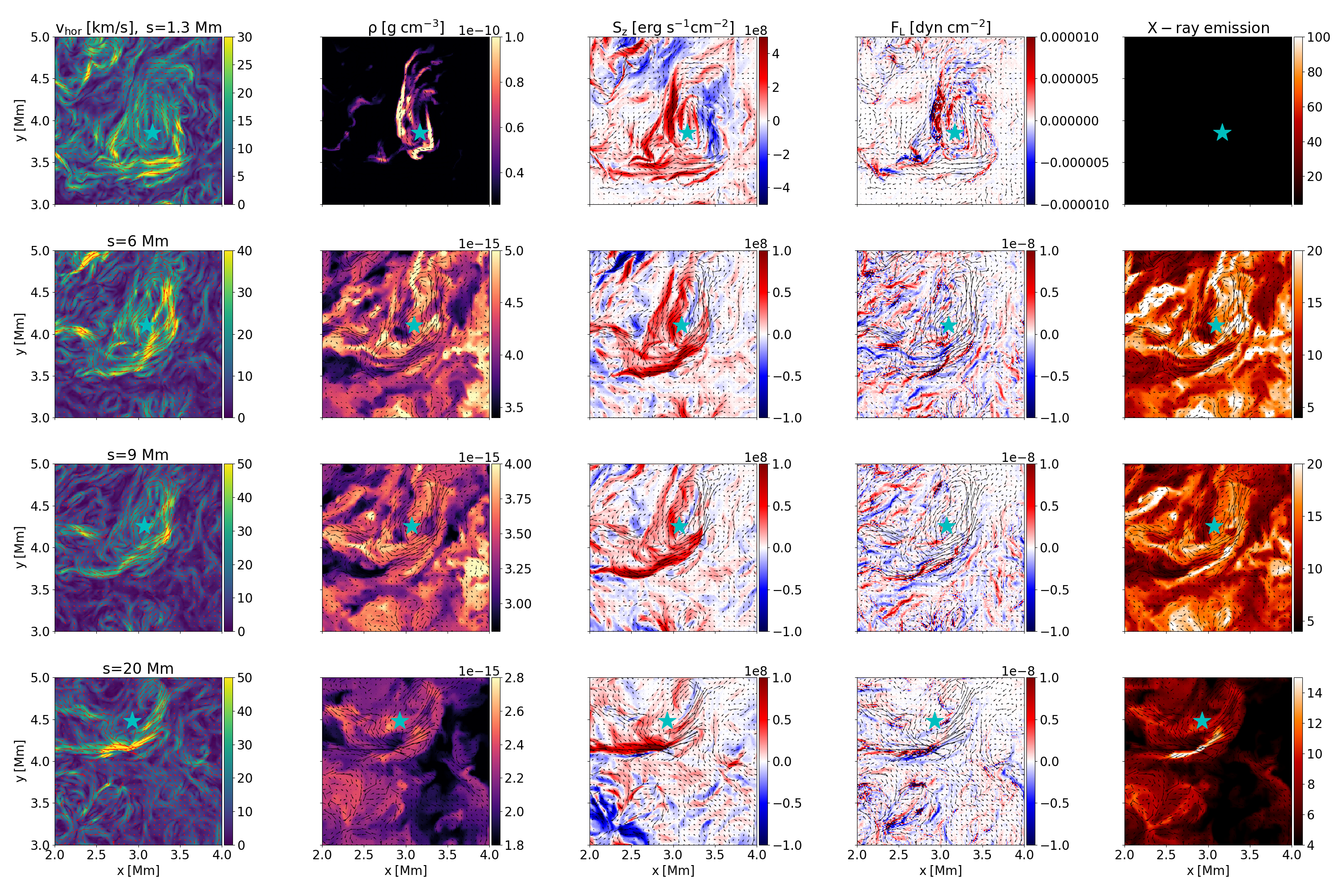}}
  \caption{Cuts perpendicular to the loop axis at different distances to the photospheric layer through the structure shown in Fig. \ref{fig:swirl_fp}. From left to right: Velocity perpendicular to the loop axis, density, axial component of the Poynting flux, axial component of the Lorentz force, X-ray emission. From the top to the bottom row, cuts are shown at values of the axial coordinate $s$ of 1.3 Mm, 6 Mm, 10 Mm and 20 Mm. The arrows show direction and magnitude of the velocity field. The units of the X-ray emission are $\rm{DN\; s^{-1} pix_{MUR}^{-1}Mm^{-1}}$. The center of the main vortex flow is marked with a star. The vortex center has been determined as the location of peak vorticity magnitude computed from the velocity field smoothed with a Gaussian with an FWHM of 500 km. For a discussion, see Sects. \ref{section:atmo_coup} and \ref{section:atmo_coup_disc}.}
  \label{fig:cuts_hor}
\end{figure*}

\begin{figure*}
\resizebox{\hsize}{!}{\includegraphics{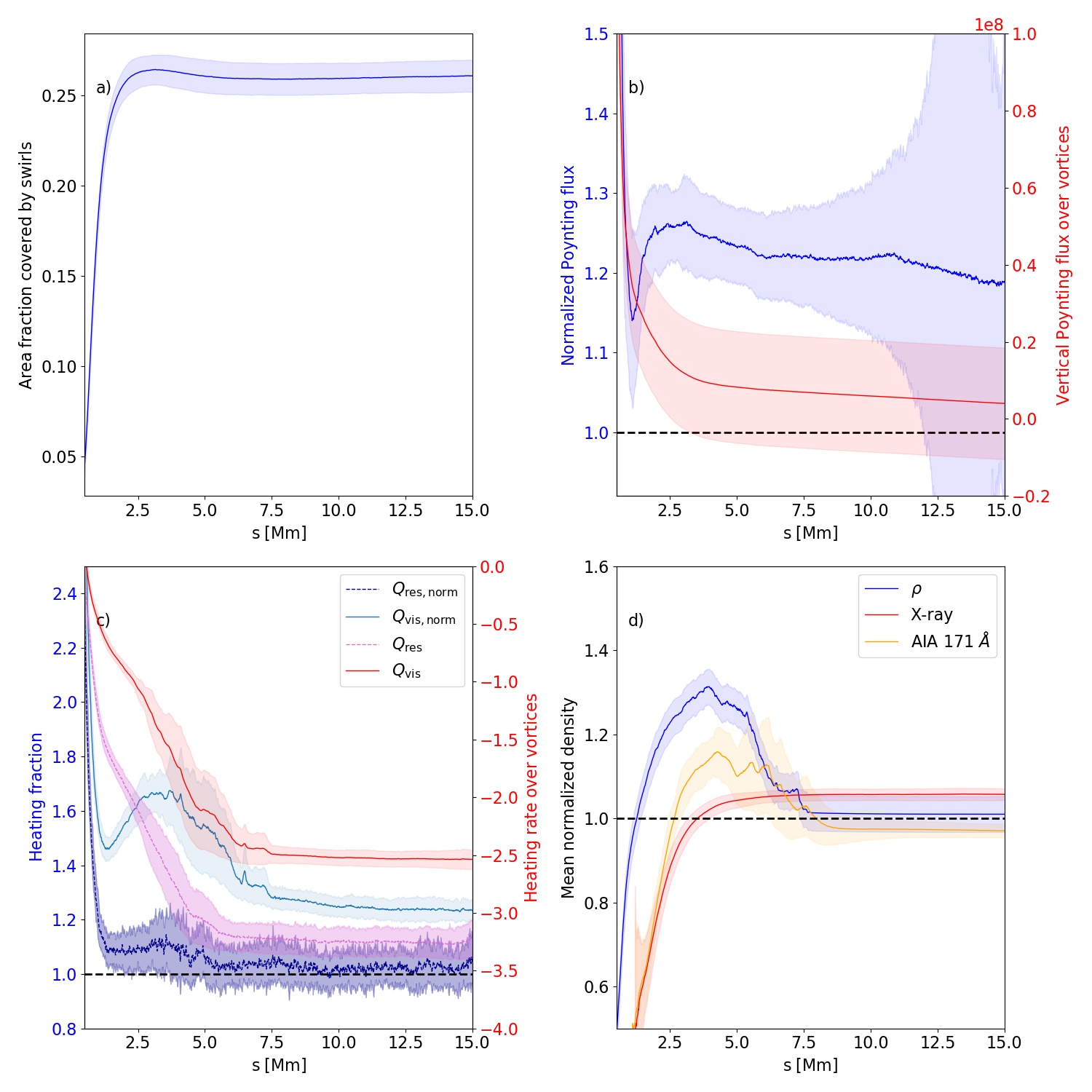}}
  \caption{Dependence of swirl properties on the axial coordinate along the loop. (a) Area fraction covered by vortices, (b) averaged Poynting flux over vortices (red) and Poynting flux over vortices normalized by the Poynting flux averaged over the loop cross section at the same distance along the loop (blue), (c) normalized viscous (green) and resistive (red) heating fraction over vortices, (d) normalized density and emission in the X-ray and the 171 \AA\ bands over vortices. The coordinate s refers to the distance to the approximate optical surface of the left loop footpoint along the semi-circular arclength of the loop. The shaded areas refer to the standard deviation of the depicted quantities at any given arclength along the loop due to variation in time. See Sects. \ref{section:swirl_prop} and \ref{section:swirl_prop_disc}.}
  \label{fig:statistics}
\end{figure*}

\section{Results}
\label{section:results_swirls}

\subsection{Atmospheric coupling}
\label{section:atmo_coup}

\begin{figure*}
\resizebox{\hsize}{!}{\includegraphics{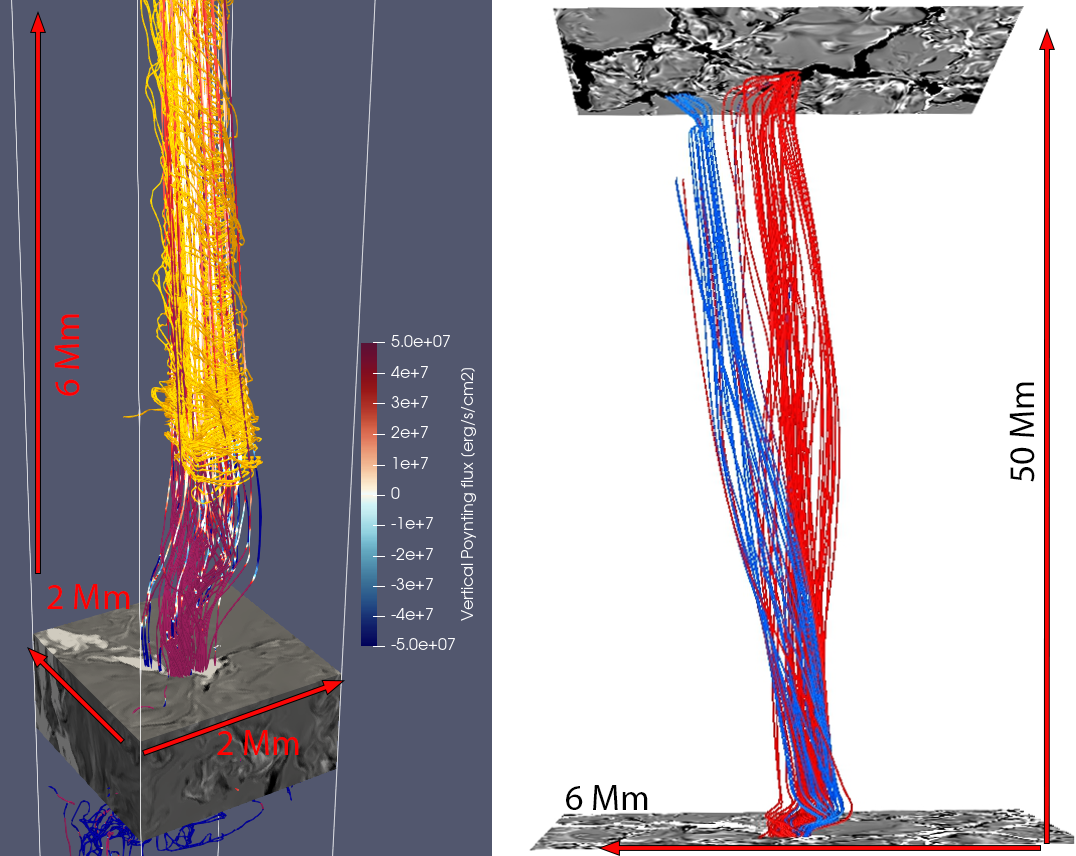}}
  \caption{3D rendering of the magnetic field lines connected to the region with enhanced swirling strength shown in Fig. \ref{fig:swirl_fp}. \textit{Left panel}: Closeup of the footpoint rooted in the swirl shown in Fig. \ref{fig:swirl_fp}. The field lines are color-coded with the axial component of the Poynting flux. Red corresponds to upward directed Poynting flux, while blue corresponds to downward directed Poynting flux. The range of the color scale is from $-5\times 10^{7}$ to $5\times 10^{7}\; \rm{erg\; cm^{2}s^{-1}}$. The vertical magnetic field is plotted on a cut at the height of the $\langle \tau \rangle = 1$ surface. The range of the color scale of the magnetic field is from -150 to 150 G.
  The orange lines illustrate the streamlines of the velocity field traced from the swirl. \textit{Right panel}: Magnetic topology of the structure rooted in the swirl. The field line bundles are colored red and blue, respectively, depending on which magnetic concentration they are rooted in at the footpoint plotted at the right. The probes show the vertical component of the magnetic field at the $\langle\tau\rangle=1$ surface at each footpoint. In the right panel, the simulation box has been compressed by a factor of five in the axial direction for better visibility. See Sect. \ref{section:atmo_coup}.}
  \label{fig:3d_flines}
\end{figure*}

\begin{figure}
\resizebox{\hsize}{!}{\includegraphics{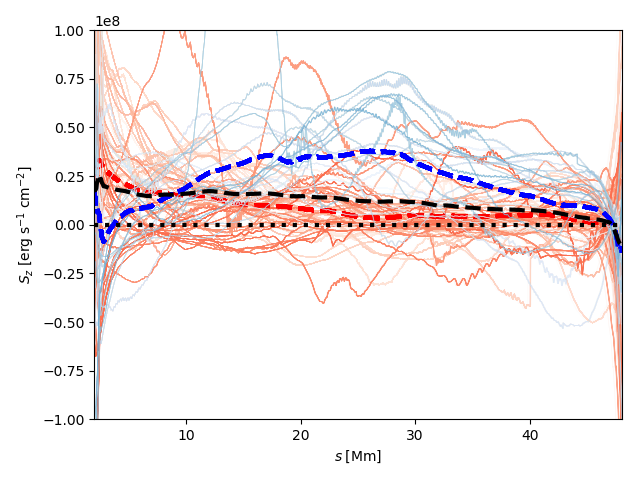}}
  \caption{Axial Poynting flux interpolated along field lines rooted in the swirling structure shown in Fig. \ref{fig:swirl_fp} traced from seed points at a height of 6 Mm. Poynting flux directed in the positive s-direction is positive, while Poynting flux directed in the negative s-direction is negative. The two populations of field lines are colored in red and blue depending on the magnetic field patch in which they are rooted at the loop footpoint at the right. The color-coding is the same as in Fig. \ref{fig:3d_flines}. The dotted black line is drawn at a constant value of zero.  The thick dashed red and blue lines show the averaged Poynting flux per flux patch, while the dashed black line shows the average taken over all traced field lines. For a discussion, see Sects. \ref{section:atmo_coup} and \ref{section:cor_heat}.}
  \label{fig:poynt_flines}
\end{figure}

Our simulations show an abundance of vortex flows in all layers of the solar atmosphere.
We have traced magnetic field lines from a prominent vortex flow in order to determine how the rotating flows in the coronal part of the simulation are magnetically connected to the solar surface. 
The swirling structure in the low corona and the footpoints of magnetic field lines connected to the structure are shown in Fig. \ref{fig:swirl_fp}.
The field lines were traced from a region of enhanced swirling strength at a distance of s=6 Mm to the solar surface (corresponding to a height of 5.86 Mm) to the photospheric layers at both footpoints of the loop. The coordinate s gives the distance to the photosphere based on the semi-circular arclength of the loop. Here we use the swirling strength computed from the smoothed velocity field. The threshold for the swirling strength was set to $0.002\; \rm{rad\; s}^{-1}$, corresponding to a rotation period of less than 50 minutes. The threshold on the swirling strength was chosen to select most of the  visually identified regions showing rotational motions.
\\
The right column shows the temperature (top) and the swirling strength computed from the smoothed flow field (bottom). 
The vertical magnetic field at the magnetic footpoints of the structure is shown in the left column. The temperature shows an enhancement along the edge of the rotating structure on the right side.\\
 At footpoint A, which is located closer to the swirl, the magnetic field lines are connected to a region displaying rotational motions in a magnetic concentration with kilogauss magnetic field strength located in the intergranular lanes. A closeup of the region corresponding to the green square in Fig. \ref{fig:swirl_fp} is depicted in Fig. \ref{fig:fp_closeup}. The bundle of field lines traced from the swirl seen in the low corona splits into two bundles of field lines at a distance along the loop arc of about 16 Mm and ends in two different magnetic concentrations with kilogauss strength at footpoint B. The region containing the two magnetic concentrations is shown in the bottom left panel of Fig. \ref{fig:swirl_fp}. The kilogauss concentrations are enclosed in pink contours. The geometry of the field lines is illustrated in Fig. \ref{fig:3d_flines}. Due to the strong guide field and consequently large Alfv\'{e}n speed, the magnetic field is only weakly twisted in the upper atmospheric layers. The field lines are twisted around each other at foopoint A where the swirl is located. While the field lines are almost vertical above the chromosphere, the streamlines of the velocity field show a helical structure.\\
At loop footpoint B, the intersections of field lines with the photosphere are less concentrated in space. The field lines are not rooted in a particular flow structure, as can be seen in the bottom left panel of Fig. \ref{fig:swirl_fp}. The concentrations the field lines are rooted in do not rotate around each other, therefore the twist in the magnetic structure must stem from footpoint A.
From 81 traced field lines, 58 field lines are rooted in the large kilogauss concentration centered at [x,y]=[3.25,4.75] Mm, while 19 field lines are rooted in the smaller concentration at [x,y]=[1.6,5.4] Mm.
 The axial Poynting flux interpolated along the two bundles of field lines is illustrated in Fig. \ref{fig:poynt_flines}. While the Poynting flux on each individual field line strongly varies, the Poynting flux averaged over all the traced field lines indicated by the dashed black line is positive through almost the whole domain.\\
The velocity and swirling strength in a plane transverse to the guide field are organized in elongated structures aligned with the magnetic field.
The velocity perpendicular to the guide field in a cut along the loop is shown in the left panel of Fig. \ref{fig:swirl_avg}. An elongated structure with increased transverse velocity can be seen with its axis located at roughly x=3 Mm. The middle panel shows the swirling strength computed from the velocity field smoothed with a Gaussian with a width of 0.5 Mm.  Small-scale swirls are abundant both inside and outside the larger rotating structure.  After smoothing, the vortex is visible as a single contiguous structure with enhanced swirling strength reaching coronal heights.  Note that for the lower effective resolution, the vortex appears to begin at a height of 1 Mm. This is due to the strong increase of the swirling strength in the corona and the horizontal expansion of structures, so that slower, smaller vortices in the chromosphere are not accurately captured. The vortex does not end at roughly 12 Mm where it disappears from the cut, but moves out of the cut at y= 4 Mm since its axis is inclined.\\
The rightmost panel of Fig. \ref{fig:swirl_avg} shows the Poynting flux averaged over a slab centered on the vortex location. The average was performed between y=3.5 Mm and y=5 Mm. The Poynting flux is enhanced at the location of the vortex. This is consistent with previous findings of enhanced Poynting flux at the locations of vortices \citep{2020ApJ...894L..17Y,2021A&A...649A.121B,2022A&A...665A.118F}. The enhancement of the Poynting flux is still present far into the coronal part of the loop. The Poynting flux stems from footpoint A at which the vortex originates and the field is twisted by the photospheric motion. There is only a weak upwards directed Poynting flux above the magnetic concentrations in which the field lines connected to the vortex are rooted in footpoint B on the opposite side.  \\

The vortex evolves with height as it propagates into the atmosphere. The height evolution of the vortex is shown in Fig. \ref{fig:cuts_hor}.
Regions of increased transverse velocity and Poynting flux appear at all four heights shown in the figure.
The vortex shows increased elongation with height and is eventually deformed into a crescent shaped flow at s=20 Mm.\\
While we find an increased upwards directed Lorentz force at the edge of the vortex in the low chromosphere, there is no clear enhancement at the vortex location in the corona. While the Lorentz force is always directed perpendicular to the magnetic field, due to the twisting of the magnetic field in the chromosphere, the magnetic field has a strong horizontal component in the lower atmosphere that allows for an upwards directed force. The Lorentz force is concentrated in many oppositely directed patches at greater heights and does not seem to lead to a large-scale acceleration of material.\\
In the chromosphere, the vortex contains a structure denser than the surrounding plasma. At coronal heights, the density contrast is much lower and the area with the highest density in the field of view is located outside the vortex. For a discussion on the relation between the swirling strength and the density distribution, see section \ref{section:swirl_prop_disc}.
Generally, the X-ray emission follows the density distribution. For the lowest height of 1.3 Mm, no X-ray emission is present since the plasma is too cool to emit in this wavelength range.
With increasing height, the X-ray emission begins to show a bright elongated structure at the edge of the vortex that becomes the brightest structure in the field of view at s=20 Mm.\\

\subsection{Swirl properties}
\label{section:swirl_prop}

To investigate general properties of swirls across a large range of atmospheric heights, we conducted a statistical study of swirl properties analogous to the study in \citet{2020ApJ...894L..17Y} and \citet{2021A&A...645A...3Y}.
To study how swirls influence quantities such as Poynting flux, heating rate, and density, we compute the average of these quantities over areas where the swirling strength exceeds a certain threshold, then normalize by dividing by the averaged quantity over the full cross-sectional domain at a certain arclength s.
In order to compare our results to \citet{2021A&A...645A...3Y}, we choose a threshold of 0.0628 $\rm{rad\; s^{-1}}$ for the swirling strength. This is equivalent to selecting only events with a rotation period shorter than $\tau=100\; \rm{s}$ if the swirls were rotating uniformly. Due to the steep decrease in density, the rms velocity and thus the swirling strength strongly increase with height in the chromosphere and transition region. We nevertheless chose a fixed value for the swirling strength threshold for simplicity since any choice of threshold would introduce a bias to the results. While the choice of threshold does affect the results, the general trend of an increased Poynting flux and dissipation rate over vortices remains the same. The average swirling strength for the small-scale swirls in the coronal part of the loop is roughly $0.05\; \rm{rad\; s^{-1}}$, vortices with a swirling strength above our threshold are thus above the average swirling strength in the loop cross-section. The profiles were averaged over 34 snapshots taken over a time range of 34.9 minutes.\\
The swirl properties as a function of height are displayed in Fig. \ref{fig:statistics}.
 The filling factor or area fraction covered by swirls as a function of height is shown in panel (a). The area coverage is low at the base of the chromosphere with less than 5 \%, reaches a maximum in the transition region with about 26 \%, then decreases slightly and remains roughly constant in the corona since the magnetic field cannot expand due to the limitations of our setup.\\
 The Poynting flux shown in Fig. \ref{fig:statistics} panel (b) is increased above vortices in the lower chromosphere. The averaged non-normalized Poynting flux over vortices decreases strongly in the transition region. The contribution from vortices strongly decreases in the upper layers of the chromosphere. After a local minimum at 1.15 Mm, it increases to a value of 1.26 to reach a local maximum at about 2.5 Mm. Subsequently, the normalized Poynting flux decreases and levels off at an enhancement of roughly 20 \% in the corona, then decreasing further with increasing height. We checked the different components of the Poynting flux and found that the Poynting flux due to perpendicular flows is increased, while the flux due to vertical motions is even decreased in the swirls compared to the average. The average net Poynting flux above vortices at s=10 Mm is $5.9\times 10^{6}\; \rm{erg\; s^{-1} cm^{-2}}$. In total, small-scale vortices contain $5.5\times 10^{23}\; \rm{erg\; s^{-1}}$ at this height. The Poynting flux over vortices at the height of the transition region is $1.6\times 10^{24}\; \rm{erg\; s^{-1}}$. \\ 
Likewise, the viscous heating shown in panel (c) of Fig. \ref{fig:statistics} is increased by about 20 \% over small-scale vortices in the corona.  The enhancement reaches values of more than 60 \% in the chromosphere and transition region after a local minimum between 1-2 Mm. The resistive heating is not enhanced in vortices above about 5 Mm. In the high Prandtl number setting that we are using, the energy dissipation occurs mostly due to viscous dissipation of flows. The total heating therefore closely follows the behavior of the viscous heating. Absolute values for the total heating rate in the corona at s=10 Mm and at the transition region are $\rm{Q_{tot}=10^{-3.2}\; erg\; cm^{-3}s^{-1}}$ and $\rm{Q_{tot}=10^{-2}\; erg\; cm^{-3}s^{-1}}$.\\
The vortices are evacuated up to a height of about 1 Mm above the photosphere, as shown in panel (d) of Fig. \ref{fig:statistics}. The density is decreased by about 50 \% in the lower chromosphere and increased in the upper chromosphere by about 30 \%. The increase in density leads to enhanced emission in the 171 \AA\ channel of AIA. The emission is increased by about 15 \% at a height of 4 Mm. In the corona, the density of the vortices does not differ significantly from the background density.\\

 At 8 Mm, the emission over vortices drops to values below the mean value. In contrast to the emission in the 171 \AA\ channel, the X-ray emission is only increased by about 5 \% in the corona compared to the average emission at a certain height. The chromosphere is too cool to emit in either of the wavelength ranges, therefore the emission drops sharply below the transition region.

\subsection{Additional events}

\begin{figure*}
\resizebox{\hsize}{!}{\includegraphics{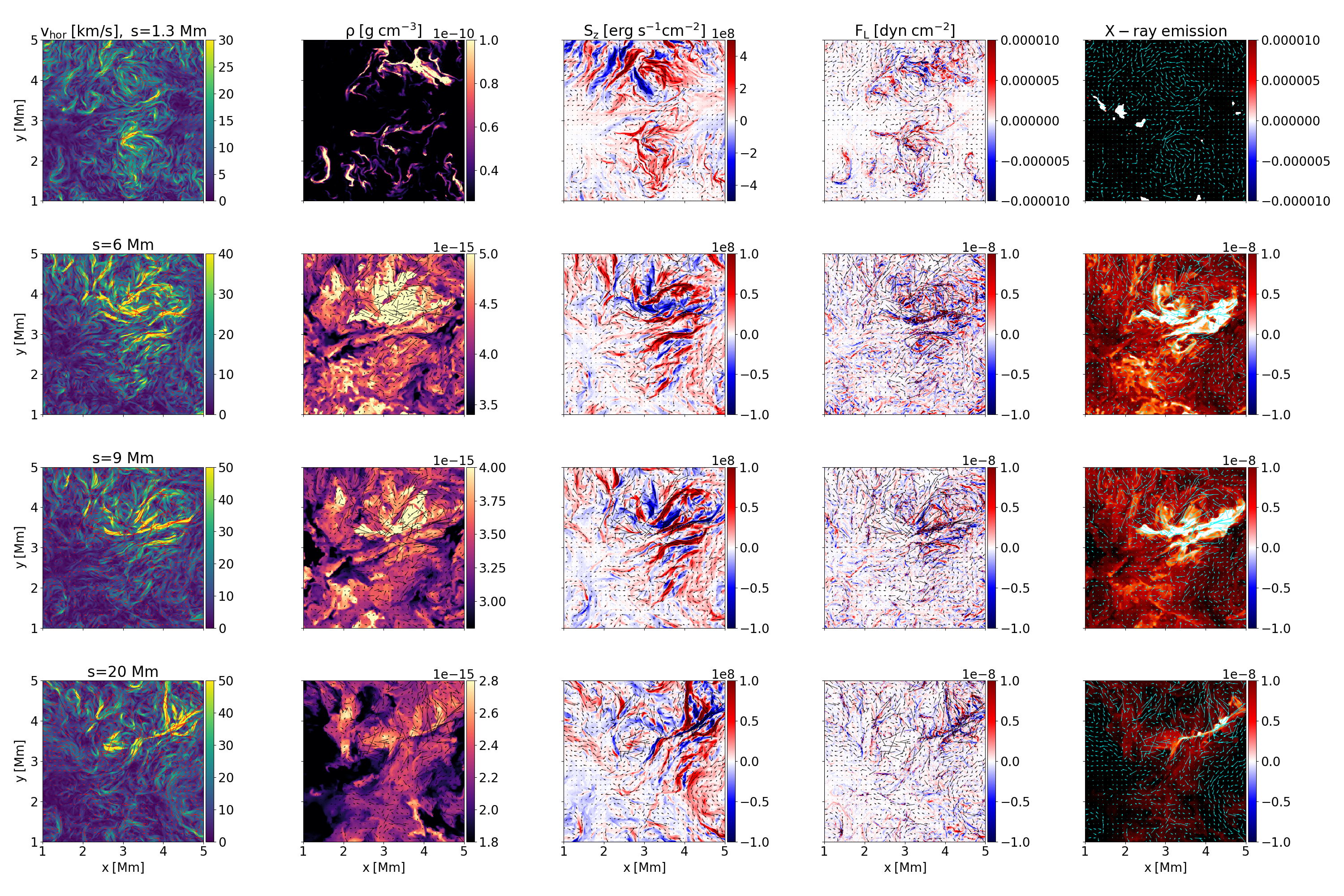}}
  \caption{Perpendicular cuts at different distances to the photospheric layer for the complex magnetic footpoint located at [x,y]=[0.5,1.5] Mm. From left to right: Transverse velocity, density, Axial component of the Poynting flux, axial component of the Lorentz force, X-ray emission. From the top to the bottom row, cuts are shown at values of the axial coordinate s of 1.3 Mm, 6 Mm, 10 Mm and 20 Mm. The arrows show direction and magnitude of the velocity field. The field of view has been centered on the footpoint assuming periodic boundary conditions. The domain has been shifted by 3 Mm in the x- and y direction. The units of the X-ray emission are $\rm{DN\; s^{-1} pix_{MUR}^{-1}Mm^{-1}}$.}
  \label{fig:cuts_appendix}
\end{figure*}

\begin{figure*}
\resizebox{\hsize}{!}{\includegraphics{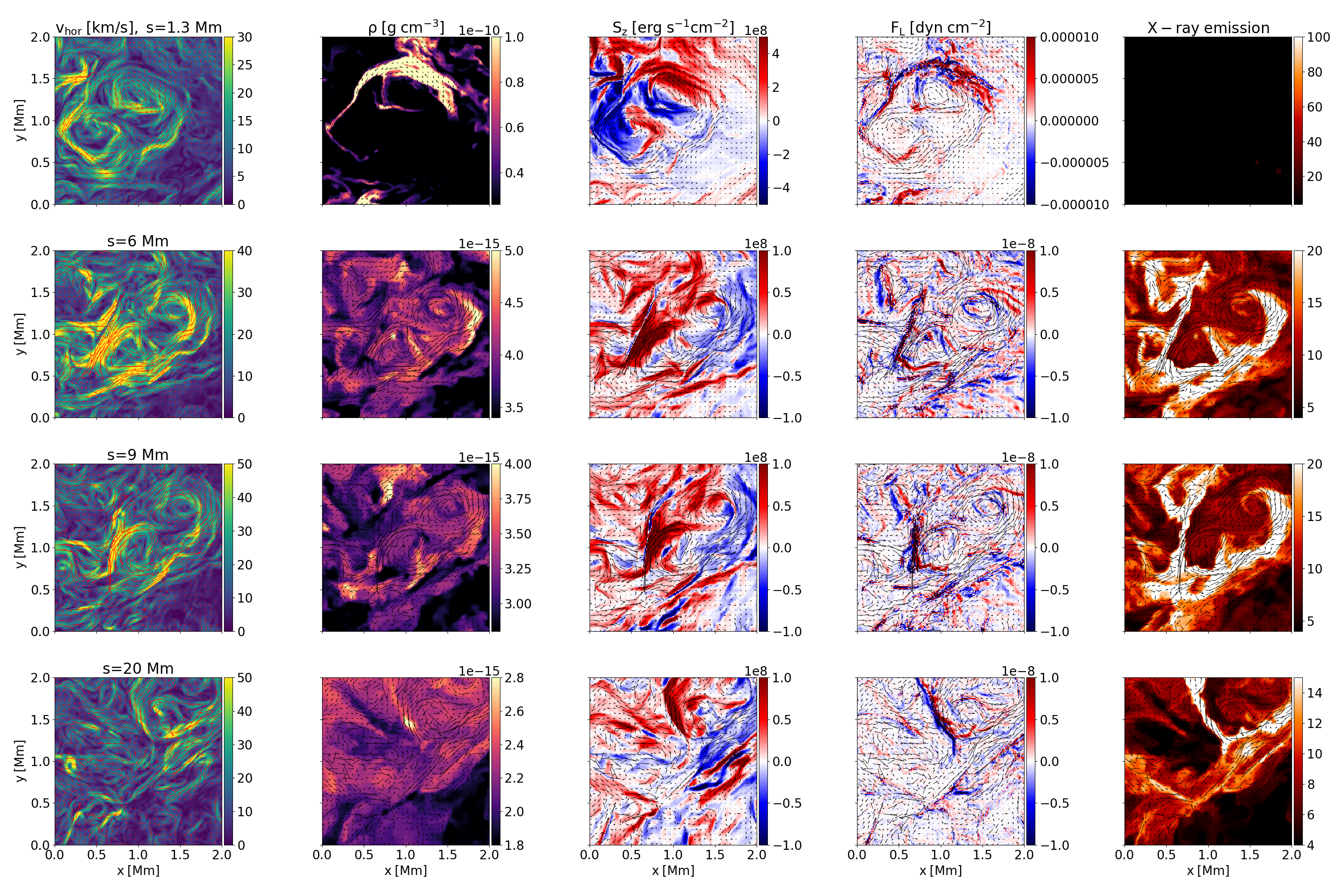}}
  \caption{Same as Fig. \ref{fig:cuts_hor}, but for a different case at a time of 5.17 min. Cuts perpendicular to the loop axis at different distances to the photospheric layer along the loop arclength. From left to right: Velocity perpendicular to the loop axis, density, axial component of the Poynting flux, axial component of the Lorentz force, X-ray emission. From the top to the bottom row, cuts are shown at values of the axial coordinate $s$ of 1.3 Mm, 6 Mm, 10 Mm and 20 Mm. The arrows show direction and magnitude of the velocity field. The units of the X-ray emission are $\rm{DN\; s^{-1} pix_{MUR}^{-1}Mm^{-1}}$.}
  \label{fig:cuts_hor2}
\end{figure*}

\begin{figure*}
\resizebox{\hsize}{!}{\includegraphics{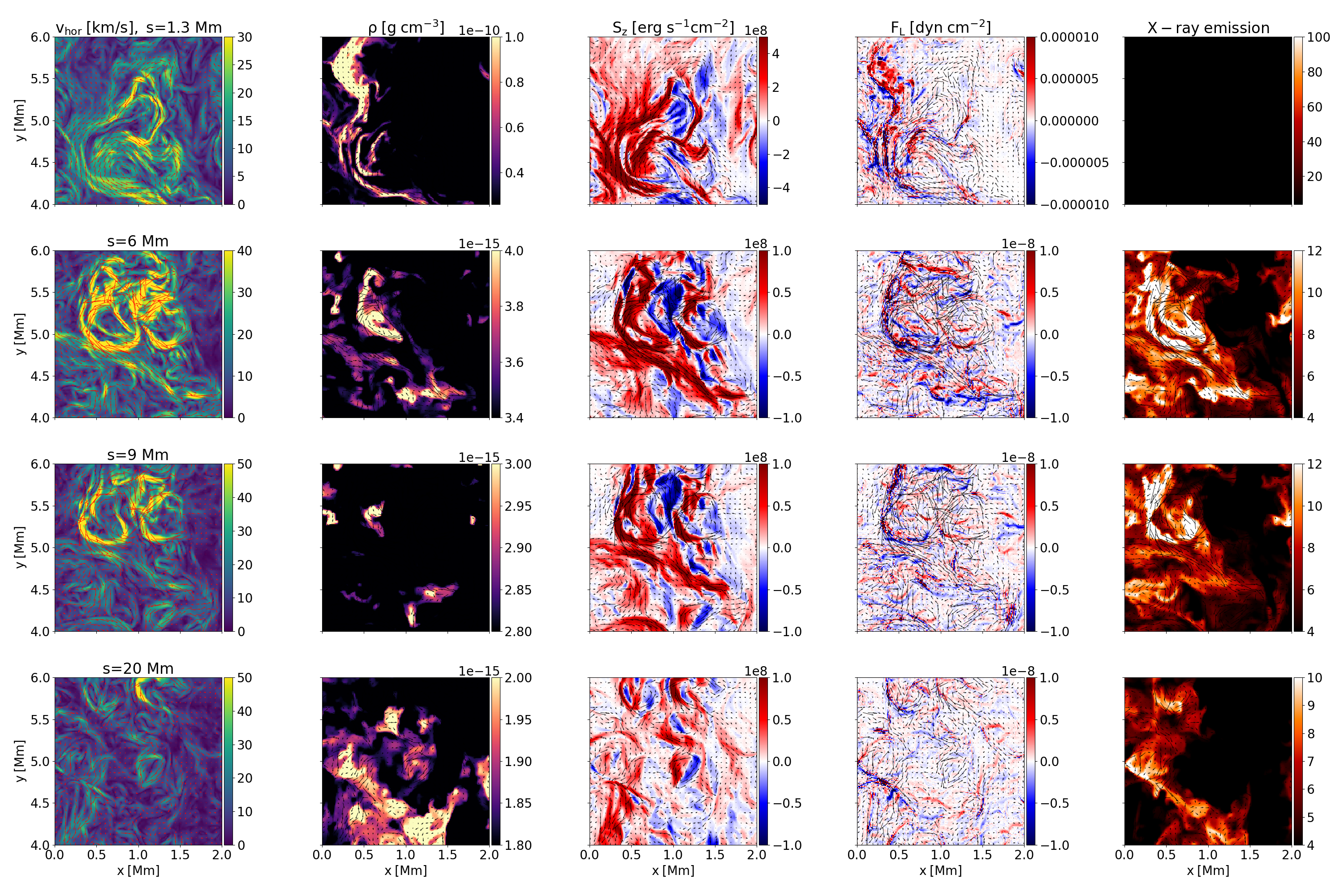}}
  \caption{Same as Fig. \ref{fig:cuts_hor}, but for a different case at a time of 13.3 min. Cuts perpendicular to the loop axis at different distances to the photospheric layer through the structure shown in Fig. \ref{fig:swirl_fp}. From left to right: Velocity perpendicular to the loop axis, density, axial component of the Poynting flux, axial component of the Lorentz force, X-ray emission. From the top to the bottom row, cuts are shown at values of the axial coordinate s of 1.3 Mm, 6 Mm, 10 Mm and 20 Mm. The arrows show direction and magnitude of the velocity field. The units of the X-ray emission are $\rm{DN\; s^{-1} pix_{MUR}^{-1}Mm^{-1}}$.}
  \label{fig:cuts_hor3}
\end{figure*}

We have identified several more swirling events that reach up into the upper chromosphere and low corona. Two such events are presented in Fig. \ref{fig:cuts_hor2} and Fig. \ref{fig:cuts_hor3}. Similar to \citet{2021A&A...649A.121B}, we frequently find events consisting of superpositions of swirls, especially with increasing distance to the photosphere. The swirl we have analyzed in this study is rooted in a relatively simple and isolated footpoint that shows a clear anticlockwise rotational motion (see Fig. \ref{fig:fp_closeup}). Most magnetic patches have a more complex structure and show a combination of rotation- and shearing motions. A superposition of events originating from a large magnetic footpoint containing several magnetic flux concentrations of kilogauss strength is shown in Fig. \ref{fig:cuts_appendix}. This footpoint is connected to the regions with the highest emissivity in X-rays within the simulation domain. While several separate rotating regions are recognizable at a height of 1.3 Mm, the flow field does not show a clear swirling pattern above that height. Instead, bright X-ray emission is present at the location of strong shear flows. 

\section{Discussion}
\label{section:discussion_swirls}

\subsection{Atmospheric coupling}
\label{section:atmo_coup_disc}

The rotating structure seen at coronal heights is magnetically connected to the photosphere and chromosphere.
Strong vortices are mainly found in regions above strong magnetic concentrations. These regions show increased transverse motions and as a consequence heightened Poynting flux, since the magnetic field communicates the photospheric motions upward due to the tension force.\\ 
Vortices twisting the magnetic field lines have been found to be important for the transport of Poynting flux in the chromosphere \citep{2020ApJ...894L..17Y}, but so far it has been unclear whether they transport Poynting flux beyond the transition region or if most of the Poynting flux is either dissipated in the chromosphere or reflected at the transition region.
We find a contiguous structure with enhanced Poynting flux extending high into the atmosphere. The Poynting flux strongly drops at a height of 1-2 Mm, but the enhancement reaches far into the coronal part of the loop. Therefore, even though dissipation and possibly reflection do take place at the transition region, a large fraction of the Poynting flux still reaches the corona.\\
The structure under consideration has one footpoint (A) rooted in the swirl and two footpoints (BI/II) on the other end of the simulation box. Poynting flux can be injected both by internal motions within magnetic flux concentrations that twist the magnetic field or by relative motions of magnetic field concentrations that lead to the magnetic flux tubes being wrapped around each other. In this case, the Poynting flux averaged over the traced magnetic field lines is positive throughout most of the length of the box, indicating that the majority of the injected Poynting flux is indeed coming from the vortical motion inside a single magnetic flux concentration. It has been a longstanding debate what the exact mechanism is that transports energy into the upper solar atmosphere and subsequently releases it. The original idea by Parker of field lines braided by random motions at the boundaries \citep{1972ApJ...174..499P,1983ApJ...264..642P} has been further developed into the fluxtube tectonics model. In this model, heating occurs at the boundaries of flux tubes that are braided by photospheric motions \citep{2002ApJ...576..533P}. In addition to the relative motion of whole flux tubes, internal motions within a magnetic flux element can drive the coronal magnetic field. Packets of Alfv\'{e}n waves that are launched by such internal motions and generate turbulence in the corona were studied by \citet{van_Ballegooijen_2011}, who named this concept "dynamic braiding". 
While an extensive study of the relative contributions of braiding of magnetic field by external or internal footpoint motions is out of the scope of this work, the rotating motions inject sufficient Poynting flux into the atmosphere that is transmitted past the transition region to heat a strand to several million Kelvin.\\
Internal footpoint motions could therefore be sufficient to heat loops of a few million Kelvin, while very hot loops might be heated by a different mechanism. Our scenario is similar to the Alfv\'{e}n pulses found by \citet{2021A&A...649A.121B} since we do not see an oscillatory motion. The swirl lasts for roughly three minutes, instead of an oscillatory motion in the photosphere, it arises from coherent internal motion within a flux element.\\\\
In addition to energy transport, vortices could play a role in the transport of mass. Distorted magnetic field lines can lead to an upwards directed Lorentz force \citep{2017ApJ...848...38I}.
In hydrodynamic equilibrium, the gravitational force acting on the plasma is balanced by the pressure gradient.
We find that the upward directed Lorentz force is of the same order as the force caused by the pressure gradient.
The resulting dense structure in the chromosphere shown in the second column of Fig. \ref{fig:cuts_hor} is roughly 500 km wide and 1 Mm long at a height of 1.3 Mm, consistent with the dark structures observed by \citet{2009A&A...507L...9W}. Jets accelerated by the Lorentz force have also been investigated in \citet{2017Sci...356.1269M}. Another possible effect that could lead to the increased density in vortices in the lower atmosphere is downflowing plasma being trapped in magnetic flux tubes.
The increase in Lorentz force is not equally distributed over the rotating structure but enhanced at the leftmost edge in the investigated example. 
\citet{2012Natur.486..505W} found that the highest Doppler shifts were found in the outer part of the ring-like structures where the centrifugal forces are largest, which is consistent with our simulations.\\
Various types of motion occur within magnetic concentrations at the photosphere. In addition to rotational motions, shearing of field lines occurs and magnetic concentrations merge and split up again. Vortices might act to transport Poynting flux into the upper atmosphere, but this does not automatically mean that the energy is dissipated. While we do find increased heating of the vortices, the relation between vortices and bright strands is more complicated. Instead of a hot, dense, and bright loop that is clearly distinct from its environment we find that the emission has a complex structure similar to the coronal veil picture studied in detail in \citet{2021arXiv210614877M}. This phenomenon has also appeared in other numerical studies, such as \citet{2005ApJ...618.1031G, 2014ApJ...795..138W, 2014ApJ...787L..22A}. In the simulation of \citet{2014ApJ...787L..22A}, strands arise from velocity sheared regions with enhanced emissivity. The vortices are generated from the Kelvin-Helmholtz instability in response to transverse oscillations of the entire magnetic flux tube. There is no one-to-one correspondence between a bright strand and a vortex. 
The simulations by \citet{2014ApJ...787L..22A} omit gravitational stratification and do not contain a chromospheric or convection zone layer. The oscillations of the loop structure are driven by a sinusoidal driver.
In contrast to this, in our simulations complex internal motions are already present in the photosphere at individual loop footpoints. It is not straightforward to determine whether individual small-scale vortices in our simulation are driven directly by rotating magnetic flux tubes or if they are created by instabilities affecting larger structures. While not all vortices found in the corona seem to be connected to photospheric counterparts, individual large-scale swirls, at least, can be traced back to corresponding rotating motions in the photosphere. This suggests that some of the vortices in our simulations are indeed driven by rotating flux tubes or motions within them and not produced directly in the corona by instabilities.
This does not rule out the possibility that such instabilities contribute to the generation of vortices in the corona or effects such as a cascade of vorticity to small scales producing vortices on various spatial scales as suggested by \citet{2021A&A...645A...3Y}.  

\subsection{Swirl properties}
\label{section:swirl_prop_disc}

Several simulations have shown that vortex flows could contribute to chromospheric heating and transport Poynting flux into the corona.
\citet{2021A&A...645A...3Y} have studied the statistical properties of vortices.\\
In agreement with their study, we find that the area fraction covered by vortices increases with height due to the expansion of the magnetic flux tubes.\\
The vortices are predominantly rooted in the intergranular lanes and thus evacuated due to the strong magnetic fields. In addition, a contribution from the dynamical pressure created by the centrifugal force to the total pressure further lowers the gas pressure inside the vortices \citep{2011A&A...533A.126M}. This is consistent with our findings. From a height of 1 Mm on, this behavior reverses and the vortices are instead overdense.
Increased density associated with vortices in the low chromosphere was also found by \citet{2021A&A...645A...3Y}, and by \citet{2012ApJ...751L..21K}, where the vortex forms a dense ring-like structure. A possible cause for this increase in density is the tangling of field lines, leading to a Lorentz force that lifts up dense plasma from the lower atmosphere, as can be seen in Fig. \ref{fig:cuts_hor}. Alfv\'{e}n waves have been shown to be able to cause density perturbations in coronal loops due to nonlinear effects \citep{2004ApJ...610..523T}.\\
An increased Poynting flux at the location of vortices due to the twisting of the magnetic field in the chromosphere has also been found by \citet{2020ApJ...894L..17Y,2021A&A...649A.121B}.
The local minimum at and the subsequent sharp rise of the Poynting flux above the transition region is possible because the signed Poynting flux is used.
At the transition region height, a larger fraction of the upward directed Poynting flux is balanced by downward directed flux, possibly from the submergence of low lying loops or reflection at the transition region.
In response to the increased influx of energy, the vortices show an increased heating rate compared to the surroundings, consistent with \citep{2021A&A...645A...3Y}.
Since the magnitude of the viscous heating depends on density, the strong increase in the heating fraction in the chromosphere is due to the increased density over chromospheric vortices. The resistive heating is also increased, but only by about 10 \% in the chromosphere while not showing an increase compared to the background at coronal heights. This is likely due to the more complex field geometry in the chromosphere compared to the corona. \citet{2021A&A...645A...3Y} find strong current sheets at the interfaces of vortices in the chromosphere and an enhancement of currents over vortices.
In \citet{2010MmSAI..81..582C}, the heating is seen mainly at the edges of magnetic flux concentrations and is Ohmic in nature. \citet{2012A&A...541A..68M} associate vortices with increased viscous dissipation and Ohmic dissipation with the edges of magnetic flux concentrations. The ratio of viscous to resistive heating in simulations depends on the magnetic Prandtl number \citep{2014ApJ...791...12B,Rempel_2016}.\\
The temperature over vortices is lower in the upper photosphere and lower chromosphere compared to the averaged temperature at the same geometrical height, consistent with the location of the vortices in the cooler intergranular lanes \citep{2012A&A...541A..68M}.
In contrast to \citet{2021A&A...645A...3Y}, who find that the temperature is increased in vortices at all heights above the base of the chromosphere, the temperature over vortices in our simulation is lower than average in the chromospheric layers and slightly increased in the corona. This could be due to the cooler material at higher densities trapped in the vortices at chromospheric heights. In the corona, the temperature is only increased by about 1-2 \% despite a 20 \% increase in the heating rate. This is not surprising if we take a look at the RTV scaling laws \citep{1978ApJ...220..643R}. The RTV scaling laws relate the maximum temperature of a coronal loop and quantities such as pressure and heating rate  under the assumption of hydrostatic equilibrium and thermal equilibrium, constant pressure and a uniform cross-section. For an increase in the heating rate of 20 \%, we would only expect an increase of the temperature by 5 \%. We find an even smaller temperature increase, but the RTV scaling laws assume a 1D atmosphere in thermal equilibrium, which is not the case in a realistic time-dependent 3D simulation.\\
In the corona, where the magnetic field is uniform, the effect of vortices on the density is negligible. The regions with the highest densities are not associated with vortex flows, despite the increased heating rate in vortices. This behavior can be understood considering the timescales involved. The density reacts to increases in the temperature with a delay. Typical axial velocities in the simulation are of the order of 50-100 km/s, therefore it takes several minutes to transport material through the loop. This is on the order of the typical lifetimes of small-scale vortices. The cooling timescale for a coronal loop is on the order of half an hour, therefore the loop remains hot after the actual heating event has ceased. The initial increase in density in the chromosphere is consistent with \citet{2021A&A...645A...3Y}, though since the upper boundary in their simulation is at a height of 2.5 Mm, they do not see the subsequent decline in overdensity of the swirls in the corona.\\
In observations, chromospheric vortices appear as dark features \citep{2009A&A...507L...9W}. This is compatible with the presence of elevated, dense and cool material that we find in the vortices in the upper photosphere and chromosphere.
Despite the increased density and heating rate for the low corona up to 7.5 Mm, the X-ray emission is only increased by about 4-5 percent.
The emission follows mainly the density distribution that is not significantly enhanced in swirls above a height of 8 Mm.
At chromospheric heights, both the emission in the 171 \AA\ channel and in X-rays is lower than the average emission at the same perpendicular slice.
 This could be due to cool material reaching larger heights in the vortices, so that the swirls are overdense in the chromosphere and low corona, but slightly cooler than the surroundings and thus too cool to be bright in X-ray emission or in the 171 \AA\ channel.\\
The emission in the 171 \AA\ channel, which has a response function that peaks around 600000 K \citep{2012SoPh..275...17L}, is increased by about 10 \% in the low corona and its behavior qualitatively follows the density distribution from a height of 2.5 Mm on. At coronal heights, the swirls are darker in the 171 \AA\ channel than the surroundings.
This is due to the high temperatures in the corona that rise well above 600000 K. The X-ray emission is enhanced in the hottest parts of the corona.\\
The brightest areas do not coincide with the highest swirling strength and the brightness contrast is small compared to other parts of the loop cross section.
Heating occurs at strong gradients in either the magnetic or velocity field. This is the case in regions with strong shear flows such as the edge of a vortex or regions where vortices interact. This is similar to \citet{2016ApJ...830...21R}, where the current first increases at the location of shear between the twisted and untwisted regions. Other types of motion, however, such as relative motion of flux tubes or pure shear motions can also lead to gradients in the velocity and magnetic field and thus to viscous and resistive dissipation.
The strand of bright X-ray emission in the right column of Fig. \ref{fig:cuts_hor} corresponds to such a dissipation region caused by the vortex flow, but since the vortex edge falls below the swirling strength threshold and the emission is not necessarily increased within the swirls themselves, this relation between vortices and bright areas is not accurately captured in the statistics.
The swirling strength criterion, when applied to the smoothed velocity field, does not capture the rotating structure in its entirety. An arbitrary threshold needs to be selected to define the vortex boundary. Different vortex identification methods should be investigated and compared.

\subsection{Implications for coronal heating}
\label{section:cor_heat}

While the original Parker model considered random motions on an infinitely conducting plate as the driver for the braiding of field lines, the model has since been extended to more realistic scenarios and the term is often used to refer to large-scale braiding of magnetic structures. Most loops, however, do not show evidence for braiding on observable scales. \cite{van_Ballegooijen_2011} instead suggest that the heating is due to motions inside magnetic concentrations.
\citet{2006A&A...459..627D} compare coronal heating due to rotational and spinning footpoint motions. In the first case, the sources of two fluxtubes are rotating around each other, entangling the fluxtubes. In the second case, the footpoints do not undergo a bulk motion, instead the field lines inside the flux tubes are tangled by internal spinning motions. The authors find that in the presence of a background field, the small-scale spinning motions are more efficient at dissipating energy for the same misalignment angle than rotational motions.\\
Our findings are more in line with the original Parker braiding model that also includes reconnection due to torsion within aligned flux tubes \citep{1982GApFD..22..195P} than with the flux tube tectonics model in which heating arises from the relative motion of flux tubes \citep{2002ApJ...576..533P}.
Earlier studies found that swirls can provide a significant amount of the energy flux to heat the chromosphere and corona \citep{2012Natur.486..505W,2021A&A...649A.121B}.
The average vertical Poynting flux at a height of 6 Mm above the swirling structure outlined by the contour in the bottom right panel of Fig. \ref{fig:swirl_fp} is $5.1\; \rm{kW m^{-2}}$. At a height of 2 Mm the average Poynting flux is $18.4\; \rm{kW m^{-2}}$ and at a height of 1 Mm above the photosphere even $127.8\; \rm{kW m^{-2}}$.\\ Large-scale swirls with a swirling strength above a threshold of $0.002\; \rm{s}^{-1}$, cover about 30 \% of the simulation domain at a height of 1 Mm, while roughly 40 \% of the Poynting flux is channeled through them. The amount of Poynting flux available at the transition region is compatible with the requirement of $10\; \rm{kW m^{-2}}$ to heat coronal plasma to several million Kelvin. \citet{2012Natur.486..505W} estimate a Poynting flux of 440 $\rm{W\; m^{-2}}$ from their numerical model contributed by vortices at the transition region height in the quiet Sun. \citet{2014ApJ...793...43C} found an energy flux of 280  $\rm{W\; m^{-2}}$ just above the transition region and 130 $\rm{W\; m^{-2}}$ in a simulation of a swirl exciting Alfv\'{e}n waves in a magnetic flux tube, while \citet{2018MNRAS.474...77M} found 300 $\rm{W\; m^{-2}}$ in a similar numerical experiment. \citet{2015Natur.522..188A} estimated the Poynting flux carried by torsional Alfv\'{e}n waves at a height of 10 Mm to be 300 $\rm{W\; m^{-2}}$. \citet{2020ApJ...894L..17Y} find the full contribution from small-scale vortices in the upper chromosphere to be 7500 $\rm{W\; m^{-2}}$. For an isolated swirling event, \citet{2021A&A...649A.121B} find 34.5 $\rm{kW\; m^{-2}}$ in the middle chromosphere.\\
Ion-neutral effects can enhance the absorption of waves in the chromosphere \citep{2012ApJ...747...87K}. The absorption of Poynting flux has been found to be enhanced in the middle and upper chromosphere in the presence of ambipolar diffusion \citep{2016ApJ...819L..11S}.\\
We find several examples of large-scale vortex flows reaching up into the corona shown in Fig. \ref{fig:cuts_hor}, Fig. \ref{fig:cuts_hor2} and Fig. \ref{fig:cuts_hor3} that are magnetically connected to photospheric vortex flows within strong magnetic concentrations, indicating that at least some of the coronal vortex flows found in the simulation are driven by footpoint motions and not generated in the corona. The generation and connectivity of vortex flows in different atmospheric layers will be the subject of future studies.\\
A significant fraction of the Poynting flux is transmitted through the transition region. In order to dissipate the injected energy, gradients in the magnetic field or in the velocity field need to form. As Fig. \ref{fig:poynt_flines} shows, the average axial Poynting flux along a set of sample field lines is positive almost through the entire length of the loop. A small fraction of the Poynting flux thus does not get dissipated and reaches the opposite transition region. The strongest dissipation occurs along the outer edge of the vortex structure along a strong velocity gradient. The heating rate is higher above the complex footpoint seen in the lower left quadrant. A superposition of swirls creates more small-scale structure that allows for dissipation. \\

\section{Conclusion}

\label{section:conclusion_swirls}

In simulations, vortices have been found both in the photosphere and in the chromosphere. These structures are not distinct, but the chromospheric vortices are rooted in their photospheric counterparts. While vortices have been observed in the low corona and it has been suggested that vortices play an important role for channeling energy and plasma into the corona \citep{2012Natur.486..505W}, the continuation of these structures beyond the transition region has never been studied in detail in simulations.\\
Using high resolution simulations of the solar surface, chromosphere, and corona, we find that vortices do not only extend into the chromosphere, but form contiguous structures that connect the photosphere with the corona.
Consistent with previous studies, vortices are energetically important especially in the upper chromosphere, showing increased Poynting flux and heating rates.
Upward acceleration of chromospheric plasma leads to a higher density at vortex locations in the chromosphere and low corona.
While vortices play an important role for energy transport and transverse density structuring of the chromosphere and low corona, their role becomes less clear with increasing height. Poynting flux and heating rate are still increased at coronal heights, but less so than in the chromosphere, and the effect on the density is small. There is a complex relationship between coronal emission and vortices. Regions with enhanced emissivity at vortex edges could potentially appear as loop strands.\\
A large variety of vortex detection methods exists, from which some, such as the $\Gamma$-functions method \citep{2001MeScT..12.1422G} or the LAVD method \citep{2016JFM...795..136H} and the Rortex criterion \citep{2022A&A...668A.118C}, allow for the identification of vortices and their boundaries without the location of the boundary being dependent on an arbitrary threshold on the vorticity or swirling strength. The Rortex criterion has been found by to be the most reliable criterion for extracting properties of the vortex such as the rotation period. For a comprehensive overview of the advantages and disadvantages of different vortex detection methods see \citet{2018A&A...618A..51T}. The influence of different criteria for the identification of vortices and their boundaries on the derived properties of vortices needs to be investigated.

\begin{acknowledgements}

This project has received funding from the European Research Council (ERC) under the European Union’s Horizon 2020 research and innovation program (grant agreement No. 695075). We gratefully acknowledge the computational resources provided by the Cobra supercomputer system of the Max Planck Computing and Data Facility (MPCDF) in Garching, Germany. This material is based upon work supported by the National Center for Atmospheric Research, which is a major facility sponsored by the National Science Foundation under Cooperative Agreement No. 1852977.

\end{acknowledgements}

\begin{appendix}
\section{Influence of the swirling strength threshold}

\begin{figure*}
\resizebox{\hsize}{!}{\includegraphics{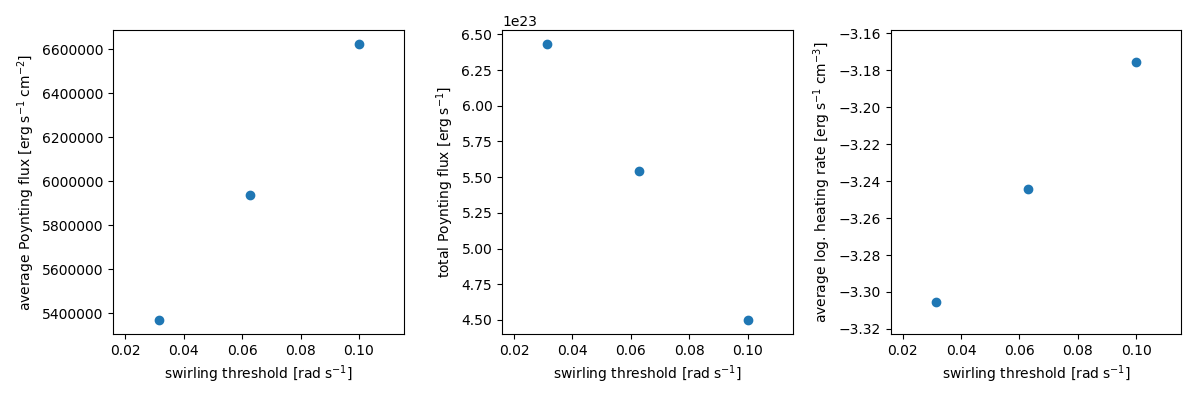}}
  \caption{From left to right: Average Poynting flux, total Poynting flux and average heating rate over vortices as a function of the swirling strength threshold at s=10 Mm. }
  \label{fig:thresh_dep}
\end{figure*}

The swirling strength criterion for the identification of vortices requires the choice of a threshold on the swirling strength to determine the location of the vortex boundaries.
In this study, events with periods shorter than 100 s are selected to compute statistical properties of vortices. The influence of the swirling strength threshold on these properties is illustrated in Fig. \ref{fig:thresh_dep}. The figure shows the average Poynting flux over vortices, the total Poynting flux over vortices summed over a cross section of the loop and the average heating rate as a function of the swirling strength threshold at a distance to the photosphere of s=10 Mm. Two different additional thresholds corresponding to periods of 468.7 s and 62.8 s are being considered.  While the threshold on the swirling strength does affect the results, the general trends remain the same. The average Poynting flux and the total heating rate consisting of the sum of the resistive and viscous heating rate are increased over vortices. 
Only including faster rotating vortices leads to a higher average Poynting flux and increased dissipation due to larger gradients in the velocity. 
While the average Poynting flux and the average heating rate over vortices increase with a higher threshold, the total Poynting flux over vortices decreases since faster rotating vortices cover a smaller fraction of the loop cross-section.

\end{appendix}

\bibliography{paper}
\bibliographystyle{aa}

\end{document}